\documentclass[%
 reprint,
 amsmath,amssymb,
 aps,
pra,
]{revtex4-2}

\usepackage{graphicx}
\usepackage{dcolumn}
\usepackage{bm}
\usepackage[utf8]{inputenc}
\usepackage[T1]{fontenc}
\usepackage{newtxtext}
\usepackage{newtxmath}
\usepackage{mathtools}
\usepackage{etoolbox}
\usepackage{braket}
\usepackage[skins]{tcolorbox}
\usepackage{tikz}
\usepackage{dcolumn}
\usepackage{subfiles}
\usepackage{pgfplots}
\usepackage{pgf}
\usepackage[export]{adjustbox}
\usetikzlibrary{calc}
\usepackage{textcomp,gensymb}
\pgfplotsset{compat=1.9}
\usepackage{placeins}
\usepackage{amsmath}
\usepackage{subcaption}
\usepackage{amssymb}
\usepackage{newtxmath}
\usepackage{ragged2e}  
\usepackage{caption}

\begin{document}


\captionsetup[subfigure]{justification=justified, singlelinecheck=false}

\makeatletter
\@ifundefined{mathdefault}{
  \newcommand{\mathdefault}[1]{#1}
}{
  \renewcommand{\mathdefault}[1]{#1}
}
\makeatother

\makeatletter
\def\@email#1#2{%
 \endgroup
 \patchcmd{\titleblock@produce}
  {\frontmatter@RRAPformat}
  {\frontmatter@RRAPformat{\produce@RRAP{*#1\href{mailto:#2}{#2}}}\frontmatter@RRAPformat}
  {}{}
}%
\makeatother

\title{Elastic scattering of twisted electrons by CO$_2$ molecules at high energies\\}
\author{Raul Sheldon Pinto}

\email{rchoubisa@pilani.bits-pilani.ac.in}
\author{Rakesh Choubisa*}%
\affiliation{Department of Physics, Birla Institute of Technology and Science, Pilani, Pilani Campus 333031 (India)
}%

\date{\today}

\date{\today}

\begin{abstract}
Elastic scattering of a twisted (Bessel) electron beam by CO$_2$ molecules is studied theoretically at high energies. The molecule's structure is optimized using coupled cluster theory and density functional theory with correlation-consistent and Pople basis sets. Coulomb potentials are used in the static approximation. The differential and total scattering cross-sections are computed in the first Born approximation. All cross-sections are orientation-averaged using a passive rotational averaging technique. The scattering is studied by the impact of the twisted beam with topological charges in the range $m_l$ = 1 and $m_l$ = 20. The cross sections are, in addition, averaged over the target's impact parameters, which accounts for the cross sections of a large distribution of CO$_2$ molecules. Finally, the molecule's total cross-section by plane waves and twisted beams is reported. The proposed methodology can be applied to study any polyatomic molecule, regardless of its structure.  
\end{abstract}

\maketitle


\twocolumngrid

\section{\label{sec:level2} Introduction \protect\\}

The recent breakthrough in realizing quantum particles with Orbital Angular Momentum (OAM) has attracted the attention of many researchers \cite{zou2023recent,molina2007twisted,grillo2015holographic,afanasev2021elastic,maxwell2021manipulating}. Such particles possessing OAM are termed twisted/vortex particles, one of them being the twisted electron.  Twisted electrons have been reported to be potential candidates in applications of fields such as high-resolution imaging \cite{grillo2016generation}, nanoparticle manipulation \cite{shi2022multislice}, and quantum information and computing \cite{larocque2018twisted}. Researchers have achieved the generation of twisted electrons with OAM up to 1000 (in atomic units) \cite{zou2023recent}. Owing to the OAM of twisted electrons, they possess additional degrees of freedom, enabling them to act as higher information carriers (qutrits, qudits, etc.). Their ability to perform multiple control operations simultaneously reduces complexity in quantum circuits and increases the efficiency of quantum algorithms, thus increasing the speed of computations \cite{wang2020qudits}. The helical motion of the twisted electrons about the propagation axis produces large magnetic fields (in the order of Teslas) along the axis, leading to extremely high magnetic field generation at nanoscales \cite{grillo2017observation}. In addition, magnetic fields can interact with the magnetic moments of targets and thus can be used to characterize the magnetic properties of targets. To realize such applications, it is primal to understand the interactions between twisted electrons and matter.

Among many possible reactions occurring during the interactions of electrons with matter, the study of elastic scattering remains one of the most powerful and important methods to investigate fundamental interactions in the quantum regime. Elastic scattering is a non-radiative interaction channel wherein there is no energy transfer between the incident projectile and the target. Although elastic scattering forbids energy transfer, the projectile might undergo a change in direction after interacting with the target. 

With the scattering and ionization studies of twisted electrons picking up pace, the cross sections of atoms and small molecules by twisted electron impacts have been reported. Ionization studies have been carried out on H and He atoms, H$_2$, H$_2$O, CH$_4$, NH$_3$ molecules \cite{dhankhar2020double,dhankhar2022triple,dhankhar2023dynamics,Plumadore_2020}. Elastic scattering of twisted electrons for atomic targets and diatomic molecular targets, with an example of a H$_2$ molecule, was reported in the previous decade \cite{kosheleva2018elastic,serbo2015scattering,maiorova2018elastic}. The work on diatomic molecules was the first reported work on twisted electron elastic scattering by molecules \cite{maiorova2018elastic}. The cross-sections were performed in the First Born approximation (FBA) with Yukawa potentials and reported only for parallel and perpendicular orientations of H$_2$.  Although many theoretical calculations are performed, there has not been much progress in the experimental measurement of atomic and molecular cross-sections by twisted electron impact. However, studies on the generation of vortex beams and preliminary/related studies on their properties have begun \cite{bu2024generation,herring2011new,grillo2014generation,grillo2017observation}. 

An effective scattering calculation requires an accurate target ground state. Particularly in the case of molecules, post-Hartree-Fock methods have been reported to be more accurate than Density Functional Theory \cite{zinoviev2017comparison}. The price to pay for the accuracy is their high computational expense. With the advancement in quantum chemistry tools such as GAUSSIAN, GAMESS, and Multiwfn, complex calculations are possible with ease. GAUSSIAN and GAMESS, through their inbuilt functionalities, enable the computation of efficient and rigorous post-Hartree-Fock methods with ease and minimal human interference. Furthermore, post-processing the output files of GAUSSIAN and GAMESS using Multiwfn is highly simplified.

Addressing the above challenges and gaps, this paper presents a theoretical method to obtain elastic cross-sections of molecules by twisted electron impact with the help of quantum chemistry tools. Following this, we compute the cross sections of the CO$_2$ molecule. CO$_2$ contains 22 electrons and 3 nuclear centers and lacks a permanent dipole moment. The study is performed at high energies in the FBA. Initially, the molecular orbitals (MO) are obtained after the Hartree-Fock stage of the Coupled-Cluster Singles and Doubles (CCSD) calculation. The electron correlations are then computed by CCSD. Following this, the interaction potential is calculated at each point in the interaction space using the electron charge density obtained after the CCSD calculation. Furthermore, the interference effects due to the multicentered nature of the molecule are also seen. The subsequent sections are discussed in the order: Section II describes the theoretical formalism for the calculation of the differential and total cross sections of molecules by the impact of twisted beams. Section III presents the computational details. Section IV describes the results and discussions of our calculations of the differential and total cross sections. We conclude this paper in Sections V and VI. The calculations of various cross sections are reported in atomic units ($m_e = \hbar = a_0 = 4\pi \epsilon_0 = 1$) unless otherwise stated.

\section{Theoretical Formalism}

\subsection{Monochromatic Bessel Beam}

One of the many representations of twisted electron beams is the Bessel electron beam. Bessel beams are the solutions to the Helmholtz equation in cylindrical coordinates \cite{khonina2020bessel}. A monochromatic Bessel electron beam consists of electrons with a definite OAM and has cylindrical symmetry. The beam propagates helically in space and time, exhibiting topological structures due to the OAM of the electrons; the beam is thus said to possess a topological charge \cite{larocque2018twisted}. The twisted beam consists of longitudinal and transverse momentum/wave vector components; the longitudinal wave vector causes the beam to propagate along the beam axis, and the transverse component spreads it outward perpendicular to the beam axis. Such beams house a line of singularity on their axis. Figure 1 illustrates the views and the intensity profile of a Bessel beam specifically: (a) depicts the lateral view of a generalized Bessel beam, (b) depicts the cross-sectional view of a generalized Bessel beam, and (c) depicts the intensity profile of a Bessel beam with topological charge 1. The spiral structures in (a) and (b) represent the tip of the wave vector as the beam propagates.

\onecolumngrid

\begin{figure*}[h]
\centering
    \begin{subfigure}{0.33\linewidth}
        \includegraphics[width=\textwidth]{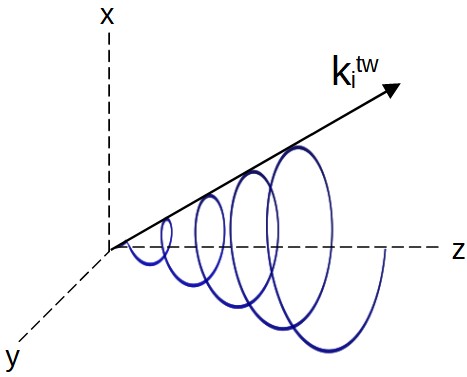}
        \caption{\justifying Lateral view of a generalized Bessel beam. The vector shows the incident wave vector.}
    \end{subfigure}   \hfill
    \begin{subfigure}{0.3\linewidth}
        \includegraphics[width=\textwidth]{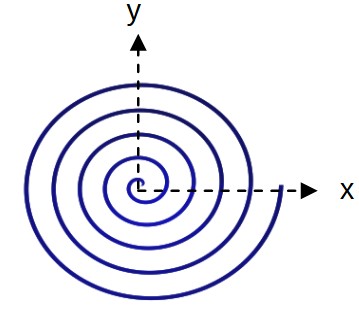}
        \caption{\justifying Transverse view of the generalized Bessel beam.}
    \end{subfigure} \hfill 
    \begin{subfigure}{0.3\linewidth}
        \includegraphics[width=0.85\textwidth]{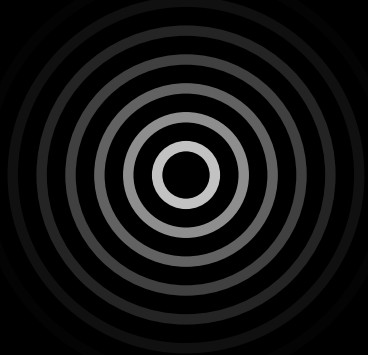}
        \caption{\justifying Transverse intensity profile of a Bessel beam with $m_l$ = 1. The axes are the same as (b).}
    \end{subfigure}
    \caption{ Spatial structure and intensity profile of the Bessel beam}
\end{figure*}

\twocolumngrid

The expression for the Bessel beam propagating in the positive z-axis described in the position representation in cylindrical coordinates is given by:

\begin{equation}
    \psi_{{k}_{i_\parallel},\varkappa, m_l}^{(t w)}\left({\rho,\varphi,z}\right)= \sqrt{\frac{\varkappa}{2 \pi}} J_{m_l}(\varkappa \rho) e^{ik_{i_{\parallel }}z} e^{im_l\varphi}
\end{equation}

where, $J_{m_l}(\varkappa \rho)$ is the Bessel function of the first kind of order $m_l$. $m_l$ is also the topological charge (OAM of the electrons). $\rho, \varphi, $ and $z$ are the standard basis in cylindrical coordinates. $\varkappa$ is the magnitude of the transverse wave vector. The wave vector of the twisted beam is denoted by $\mathbf{k_i^{tw}}$ and can be resolved into parallel and perpendicular components, given as (also depicted in Figure 2):

\begin{equation}
    \mathbf{k_i^{tw}} = \textbf{k}_{i_\parallel} + \textbf{k}_{i_\perp}
\end{equation}

where $\mathbf{k_i}_{_\parallel}$ and $\mathbf{k_i}_{_\perp}$ are the longitudinal and transverse wave vectors, respectively. They are expressed with their Cartesian components as:

\begin{equation}
\mathbf{k}_{i_\parallel}=k_{i_{\parallel}} \hat{z}=\left(k_{i} \cos \theta_{p}\right) \hat{z}
\end{equation}
\begin{equation}
\mathbf{k}_{i_{\perp}}=     \left(k_{i} \sin \theta_{p} \cos \phi_{p}\right) \hat{x} +\left(k_{i} \sin \theta_{p} \sin \phi_{p}\right) \hat{y}
\end{equation}

Here $k_i$ is the magnitude of the incident wave vector, $\theta_p$ and $\phi_p$ are the polar and azimuthal angles of $\mathbf{{k}_i^{tw}}$ respectively which are also seen in Figure 2. $e^{ik_{i_{\parallel}}z}$ is the twisted wave component propagating in the positive z direction. The phase factor $e^{im_l\varphi}$ is responsible for the twist. The magnitude of the transverse wave vector is given by $\varkappa = \sqrt{2E_i-k_{i_{\parallel}}^2} = k_i \sin{\theta_p}$. 

We now decompose Equation (1) into a superposition of plane waves, each with a definite wave vector (the wave vector is also the momentum vector), described in the momentum representation. The resultant beam is known as the generalized Bessel beam, which is given as \cite{PhysRevA.95.032703}:

\begin{figure}[h]
    \includegraphics[width=\linewidth]{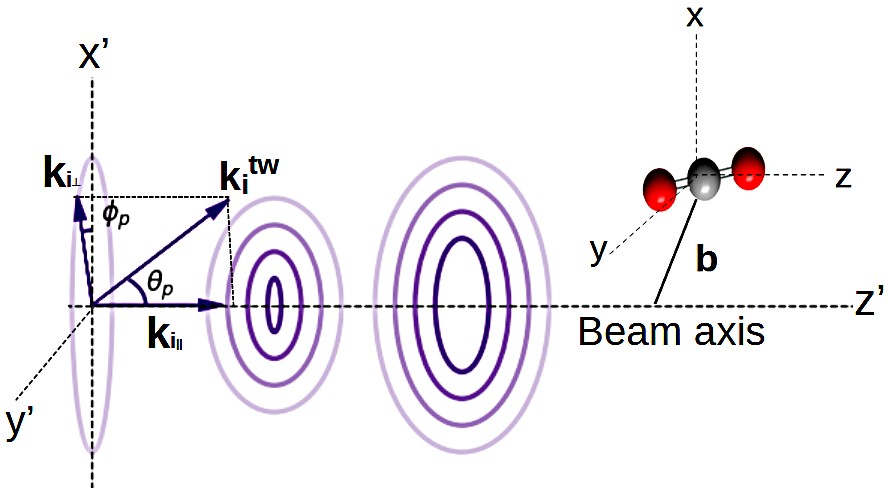}
    \caption{\justifying Bessel beam with wave vector $\mathbf{k_i^{tw}}$ with projections $\mathbf{{k}_{i_{\parallel}}}$ and $\mathbf{{k}_{i_{\perp}}}$ impacting a CO$_2$ molecule with an impact parameter $\mathbf{b}$. The projection of the beam is shown in the x'-y' plane, $\theta_p$ and $\phi_p$ are the opening angle and the azimuthal angle of the beam, respectively.}
    \label{fig:enter-label}
\end{figure}

\begin{equation}
    \psi_{\varkappa m_l}^{(t w)}(\boldsymbol{r})=\int \frac{d^2\mathbf{k_{i_\perp}}}{{(2 \pi)}^2} a_{\varkappa m_l}\left(\mathbf{k_{i_\perp}}\right) \exp(i\mathbf{k_{i}} \cdot \boldsymbol{r})
\end{equation}
    
Each plane wave $e^{i \mathbf{k_{i}} \cdot \boldsymbol{r}}$ has an amplitude:

\begin{equation}
    a_{\varkappa m_l}\left(\mathbf{k_{i_\perp}}\right)=\frac{1}{\sqrt{2\pi\varkappa}}(-i)^{m_l} \mathrm{e}^{i m_l \phi_p} \delta\left(|\mathbf{k_{i_\perp}}|-\varkappa\right)
\end{equation}

The Dirac delta function in Equation (6) picks out the value $|\mathbf{k_{i_\perp}}| = \varkappa$ in Equation (5). The detailed derivation of Equation (5) and its relation to Equation (1) is given in \cite{serbo2015scattering}.

Figure 2 illustrates the incidence of a Bessel beam on a CO$_2$ molecule with an impact parameter $\textbf{b}$. $\phi_p$ is measured on the x'-y' plane from the x' axis. The primed axes are those of the beam, and the unprimed ones are those of the molecule.

\subsection{Target Molecules}
An effective electron molecular scattering calculation requires that we have, a priori, an accurate molecular ground state. Since molecules are polyatomic, the computation of the MOs is analytically formidable. Theoretically, one may use approximation methods such as Density Functional Theory (DFT) or Hartree-Fock to obtain the MO. DFT is a density-based approach that uses electron density functionals to approximate electron correlations and other properties of the system. An overview of DFT can be found in \cite{burke2013dft}. The Coupled-Cluster Singles and Doubles (CCSD) calculation is a post-Hartree-Fock method composed of a Hartree-Fock calculation where the MOs are generated, followed by a CCSD calculation to model electron correlations.

CCSD has been proven to be one of the most accurate methods to obtain the ground state of molecules \cite{shee2019achieving}. Many studies have used CCSD to obtain the ground state \cite{arulmozhiraja2008ccsd,xiao2010elastic,moreno2019ab}. CCSD uses the single and double excitation operators of the cluster operator $\hat{T}$ (shown below) on the Hartree-Fock state (the Slater determinant), $\ket{\Phi_0} $ \cite{gallo2021periodic,zhang2019coupled} : 

\begin{equation}
    \ket{\Psi_{ccsd}} =  e^{\hat{T}} \ket{\Phi_0} 
\end{equation}

\begin{equation}
    \hat{T}= \sum_{a,i} t_i^a \hat{a}^\dagger_a \hat{a}_i + \frac{1}{4} \sum_{a,b,i,j} t^{ab}_{ij} \hat{a}^\dagger_a \hat{a}_b^\dagger \hat{a}_j \hat{a}_i
\end{equation}

$\ket{\Psi_{ccsd}}$ is the CCSD state that includes electron correlations. \textit{$t^a_i$} and $t_{ij}^{ab}$ are coefficients that can be obtained by projecting $\hat{T}$ on $\ket{\Phi_0}$. The coefficients need to be evaluated. $\hat{a}$ and $\hat{a}^\dagger$ are the annihilation and creation operators, respectively. The indices $a,b$ refer to the indices of the virtual states, whereas $i,j$ refer to those of the occupied states. 

The similarity transformed Hamiltonian is $\bar{H} = e^{-\hat{T}}H e^{\hat{T}}$. The goal is to solve the following coupled nonlinear equations iteratively that yield the coefficients \textit{$t^a_i$} and $t_{ij}^{ab}$:

\begin{equation}
    E_{ccsd} = \braket{\Phi_0|\bar{H}|\Phi_0}
\end{equation}

\begin{equation}
    \braket{\Phi_0|\hat{a}_i^\dagger \hat{a}_a \bar{H}|\Phi_0} = 0
\end{equation}

\begin{equation}
    \braket{\Phi_0|\hat{a}_i^\dagger \hat{a}_j^\dagger \hat{a}_b \hat{a}_a \bar{H}|\Phi_0} = 0
\end{equation}
$E_{ccsd}$ is the electron correlation energy. Once, \textit{$t^a_i$} and $t_{ij}^{ab}$ are obtained the computation of $\ket{\Psi_{ccsd}}$ is straightforward. 

Note that the excitations used in CCSD are only used to compute the ground-state wavefunctions and are not the transitions induced by the incident electron beam during scattering.

\subsection{Interaction potential}

The expressions for the initial- and the final-state wavefunctions for the twisted electron impact elastic scattering, respectively, are written as:

\begin{equation}
    \braket{\boldsymbol{r}|\psi_i} = \psi_{\varkappa m_l}^{(t w)}(\boldsymbol{r})  \psi_{m} (\boldsymbol{r})
\end{equation}
\begin{equation}
    \braket{\boldsymbol{r}|\psi_f} = \frac{1}{({2\pi})^{3/2}} exp(i\boldsymbol{k_s}.\boldsymbol{r}) \psi_n (\boldsymbol{r})
\end{equation}

$\psi_{\varkappa m_l}^{(t w)}(\boldsymbol{r})$ is the incident twisted electron wavefunction. $\psi_m(r)$ and $\psi_n(r)$ are the normalized initial and final CCSD molecular wavefunctions. We consider the scattered wave to be a plane wave, assuming that plane wave detectors are used that are sensitive to the wave vector \textbf{k} only. In addition, the scattered wave does not have a definite OAM, since the projectile could have imparted its OAM to the target without an energy change, which also accounts for elastic scattering. Plane waves have been used for the scattered waves previously \cite{PhysRevLett.108.074802,PhysRevA.86.023816,harris2023projectile}. The transfer of OAM is not the aim of this paper and could be studied elsewhere.

The Coulomb direct interaction potential between the incident projectile and target molecule is: 
\begin{equation}
    V_{d} = -\sum_{j=1}^M \frac{Z_j}{|\boldsymbol{r_{0}}-\boldsymbol{l_j|}} + \sum_{i=1}^{N} \frac{1}{|\boldsymbol{r_{0}}-\boldsymbol{r_i|}}
\end{equation}
Here, \textit{j} is summed over the \textit{M} nuclear centers, each having a charge $Z_j$ and position vector $\boldsymbol{l_j}$, producing an attractive potential. \textit{i} is summed over the \textit{N} molecular electrons, each having charge -1 and position vector $\boldsymbol{r_i}$, producing a repulsive potential. $\boldsymbol{r_0}$ is the incident projectile coordinate. Since we work at sufficiently high energies, the exchange and polarization effects are negligible, and we ignore their contributions. For CO$_2$, \textit{M} = 3 and \textit{N} = 22.

\subsection{Momentum Transfer}
The magnitude of the wave vector for the impact of the plane / twisted wave for the elastic scattering is $k_i = k_s = k$. The momentum transfer vector for the scattering process by plane waves is given by $\boldsymbol{\Delta} = \boldsymbol{\Delta^{pw}} = \textbf{k}_i - \textbf{k}_s$. The magnitude of $\boldsymbol{\Delta}$ in terms of the scattering angle $\theta_s$ as:
\begin{equation}
    \Delta^{pw} = \Delta =  2 k \sin(\theta_s/2)
\end{equation}
For the twisted (Bessel) wave, the momentum transfer vector of the scattering process is given by $\boldsymbol{\Delta^{tw}} = {\textbf{k}}_i^{tw} - {\textbf{k}}_s$. Its magnitude in terms of the scattering angle (twisted) $\theta_{ps}$ and the magnitude of the wave vector $k$ is:

\begin{equation}
    \Delta^{tw} = 2 k \sin(\theta_{ps}/2)
\end{equation}
$\theta_{ps}$ is the angle between $\boldsymbol{k_i^{tw}}$ and $\boldsymbol{k_s}$. In spherical coordinates, it can be written as:

\begin{equation}
\theta_{ps} = \cos^{-1}\left[\cos{\theta_{p}}\cos{\theta_{s}} + \sin{\theta_{p}}\sin{\theta_{s}} \cos{\phi_p}\right]
\end{equation}

\subsection{Scattering Amplitudes}
The elastic scattering amplitudes as functions of initial state, final state, and direct potential $V_d$ (Equation (14)) in FBA is:

\begin{equation}
    T_{fi} = \braket{\psi_f | V_d | \psi_i} 
\end{equation}

$\ket{\psi_i}$ and $\ket{\psi_f}$ are composite initial and final states (see Equation (13) for plane wave and Equation (12) for twisted wave). We henceforth call them $\ket{i}$ and $\ket{f}$ respectively: 

\begin{gather}
     T^{pw/tw}_{fi} = V_{fi} = \braket{f|V_d|i} = 
     -  \sum_{j=1}^M Z_j \int  \frac{\mathrm{e}^{i \boldsymbol{\Delta^{pw/tw}} \cdot \boldsymbol{r_0}}d\boldsymbol{r_0} }{|\boldsymbol{r_0} - \boldsymbol{l_j}|} \nonumber  \\ 
     + \int \int \frac{\rho(\boldsymbol{r}) \mathrm{e}^{i \boldsymbol{\Delta^{pw/tw}} \cdot \mathbf{r_0}}d\boldsymbol{r_0}d \boldsymbol{ r}}{|\boldsymbol{r_0} - \boldsymbol{r}|}
\end{gather} 

$\boldsymbol{r}$ is the position vector of the molecular electrons. The electron charge density $\rho(\boldsymbol{r})$ of the molecule as a function of spatial coordinates is given in terms of the inner products of the molecular wavefunctions $\psi_n$ and $\psi_m$. We assume that the molecular wavefunctions remain unchanged during the scattering process, hence $m=n$ and therefore:

\begin{equation}
     \rho(\boldsymbol{r}) = |\psi_m(\boldsymbol{r})|^2 = \sum_i \eta_i\left|\sum_\mu C_{\mu, i} \beta_\mu(\boldsymbol{r})\right|^2
\end{equation}

$\rho(r)$ is the electron density equal to the square modulus of the wavefunction, which is the sum of the products of the orbitals multiplied by their occupation number $\eta_i$; $i$ is the index of the orbitals. The orbitals are given by the square modulus of the sum over $\mu$ of the orbital coefficients $C_{\mu,i}$ times the basis function $\beta_{\mu}$(r). $\mu$ is the index of the basis function of the orbitals and $C_{\mu,i}$ corresponds to the coefficient of the $i^{th}$ orbital and $\mu^{th}$ basis function.

\subsubsection{Plane-wave scattering amplitude}
On simplifying Equation (19) using Bethe's integral given in \cite{bransden2003physics,tweed1992double}, we obtain the plane wave scattering amplitude as: 
\begin{equation}
    T_{ii}^{pw} (\boldsymbol{\Delta})  =  - \frac{2}{\Delta^2} (\alpha - \chi)
\end{equation}
Since we consider elastic scattering there is no change in the initial state of the molecule due to scattering. Hence, we represent the initial and final plane wave states by $\ket{i}$. $\alpha$ is the contribution of the attractive potential experienced by the incoming electrons due to the interaction with $M$ nuclear centers. $\alpha$ in terms of the $j^{th}$ nuclear center with position vector $\boldsymbol{l_j}$ is given by: 

\begin{equation}
    \alpha = \sum_{j=1}^{M} Z_j exp(i \boldsymbol{\Delta \cdot l_j})
\end{equation}

$\alpha$ is the contribution due to the nuclear centers. $Z_j$ is the positive charge of the $j^{th}$ nucleus. The rightmost double integral of Equation (19) simplifies to $\chi$, the target's elastic scattering form factor given by \cite{bransden2003physics}:
\begin{equation}
    \chi = \int exp(i \boldsymbol{\Delta \cdot r}) \rho(\boldsymbol{r}) d\boldsymbol{r}
\end{equation}
The electron density $\rho(\boldsymbol{r}) $ as a function of spatial coordinates is introduced into the computation from the output obtained from Multiwfn. The contribution of all electrons and the nuclei to the scattering is incorporated by input of the electron density.

To evaluate $\boldsymbol{\Delta \cdot l_j}$ in Equation (22), the angles between $\boldsymbol{\Delta}$ and $\boldsymbol{l_j}$ with fixed coordinate axes have to be calculated. $\theta_\Delta$ and  $\phi_\Delta$ are the angles that $\boldsymbol{\Delta}$ makes with the coordinate axes.  When the incident beam is along the z' axis, $\theta_\Delta$ and $\phi_\Delta$ are given by:

\begin{gather}
    \theta_{\Delta} =  \cos^{-1}\left({\frac{k}{\Delta}(1-\cos{\theta_s})}\right)
\end{gather}
\begin{equation}
    \phi_\Delta = 0 
\end{equation}

The cosine of the angle between $\boldsymbol{\Delta}$ and $\boldsymbol{r}$ to be used in Equation (23) is given by:

\begin{gather}
    \cos(\theta_{\Delta r})= \cos\theta_{\Delta} \cos\theta + \sin \theta_\Delta \sin \theta \cos(\phi - \phi_\Delta)
\end{gather}

The angles $\boldsymbol{l_j}$ make with the coordinate axes to be used in Equation (22) depend on the position vectors of the nuclei of the molecule. 
In spherical coordinates, for the initial orientation (axis of CO$_2$ parallel to z axis) Equation (22) reads:

\begin{gather}
    \alpha = 8 \exp(il_1 \Delta \cos\theta_\Delta) + 8 \exp(-il_1 \Delta
    \cos\theta_\Delta) + 6
\end{gather}

which simplifies to:
\begin{equation}
    \alpha = 16\left[cos(l_1\Delta cos(\theta_\Delta))\right] + 6
\end{equation}

$l_1$ is the modulus of the magnitudes of the position vectors of the oxygen nuclei (which was obtained from CCSD calculation to be 2.185 $a_0$).

\subsubsection{Bessel scattering amplitude}
The Bessel transition amplitude/scattering amplitude given as functions of the plane wave transition amplitude/scattering amplitude can be deduced from Equations (12), (13), (18), and (19) to be:

\begin{equation}
    T_{f i}^{t w}(\varkappa, \boldsymbol{\Delta^{tw}})=\int \frac{d^2 \mathbf{k_{i_{\perp}}}}{(2 \pi)^2} a_{\varkappa m_l}\left(\mathbf{k_{i_\perp}}\right) T_{i i}^{p w}(\boldsymbol{\Delta^{tw}}) \\
\end{equation}

Note that the notation $T_{f i}^{t w}$ is used for the twisted scattering amplitude since the scattered wave used is the plane wave, which is not the same as the incident wave which is twisted (see Equations (12) and (13)), whereas $T_{i i}^{p w}$ is used for the plane wave scattering amplitude since both the incident and scattered waves are plane waves. Also, $T_{i i}^{p w}$ in the integrand is a function of $\boldsymbol{\Delta^{tw}}$. 

The angles $\boldsymbol{\Delta^{tw}}$ makes with the fixed coordinate axes are:

\begin{equation}
    \theta_{\Delta^{tw}} =  \cos^{-1}\left({\frac{k}{\Delta^{tw}}(\cos\theta_p-cos\theta_s)}\right)
\end{equation}

\begin{equation}
\phi_{\Delta^{tw}} = \cos^{-1} \left[\frac{k( \sin\theta_p\cos\phi_p-\sin\theta_s)}{\Delta^{tw}\sin\theta_{\Delta^{tw}}}\right]
\end{equation}

For the twisted case, the cosine of the angle between $\boldsymbol{\Delta^{tw}}$ and $\boldsymbol{r}$ is given by:

\begin{gather}
    \cos\theta_{\Delta^{tw} r}= \cos{\theta_{\Delta^{tw}}}\cos {\theta} + \sin(\theta_{\Delta^{tw}}) 
    \sin \theta \nonumber\\
    \cos(\phi - \phi_{\Delta^{tw}})
\end{gather}

\subsection{Differential cross-sections}

The plane wave differential cross-section is given by:
\begin{equation}
    \left(\frac{d\sigma}{d\Omega}\right)^{pw} = |T_{ii}^{pw} (\boldsymbol{\Delta})|^2
\end{equation}

The twisted wave differential cross-section for particular values of $m_l$ is given by:

\begin{equation}
    \left(\frac{d\sigma}{d\Omega}\right)^{tw} = |T_{fi}^{tw} (\varkappa, \boldsymbol{\Delta^{tw}})|^2
\end{equation}

We also average the target molecules' differential cross-section over impact parameters. This averaging simulates a situation where the incident twisted beam is scattered by a large number of target molecules  (this mimics the macroscopic target). The cross sections of such nature are usually termed as scattering by macroscopic targets and are particularly important in the case of twisted beams \cite{serbo2015scattering}. Such cross sections are far easier to obtain experimentally for the twisted scattering case, since the impact parameters are averaged. Otherwise, one needs to place the molecule at a particular impact parameter to obtain cross sections, which is very difficult experimentally. In this case, the DCS does not depend on the topological charge of the beam but only on its opening angle $\theta_p$ \cite{serbo2015scattering,PhysRevA.95.032703}.

\begin{equation}
    \left({\frac{d \sigma}{ d \Omega}}\right)_{a v}^{tw} = \frac{1}{2 \pi \cos \theta_{p}} \int_{0}^{2 \pi} d \phi_{p} \left(\frac{d \sigma(\boldsymbol{\Delta^{tw}} (\phi_p))}{ d \Omega}\right)^{pw}
\end{equation}

\subsection{Orientation Averaging}
The preparation of a molecular orientation along a specific direction for scattering is experimentally difficult. The molecules in a typical experimental setup will be randomly oriented, which implies that only orientation-averaged DCS and TCS are useful from an experimental perspective. To average over molecular orientations, we perform a rotational averaging by rotating the scattering plane containing $\mathbf{k_i}$, $\mathbf{k_s}$, and $\mathbf{\Delta}$ vectors using the standard rotation matrix (given below), over Euler angles keeping the molecular orientation fixed. The rotation of the scattering plane, keeping the molecular orientation fixed, gives the same effect as though the molecule were rotated while the scattering plane is kept fixed. The DCS for each orientation is calculated and averaged (summed and divided by 8$\pi^2$) for a particular scattering angle $\theta_s$. The rotation matrix as a function of the Euler angles in the XYZ order is given by \cite{10019271}:

\begin{align}
\mathbf{R}_{{Euler}} & =\mathbf{R}_x(\alpha) \mathbf{R}_y(\beta) \mathbf{R}_z(\gamma) \\ \nonumber
& =\left[\begin{array}{ccc}
c_\beta c_\gamma & -c_\beta s_\gamma & s_\beta \\
c_\alpha s_\gamma+s_\alpha s_\beta c_\gamma & c_\alpha c_\gamma-s_\alpha s_\beta s_\gamma & -s_\alpha c_\beta \\
s_\alpha s_\gamma-c_\alpha s_\beta c_\gamma & s_\alpha c_\gamma+c_\alpha s_\beta s_\gamma & c_\alpha c_\beta
\end{array}\right]
\end{align}

$c_i$ and $s_i$ represent $\sin{i}$ and $\cos{i}$, respectively where $i \in \{\alpha,\beta,\gamma\}$ and $\alpha = [0,2\pi], \beta = [0,\pi], \gamma = [0,2\pi]$.

The transformed angles that $\boldsymbol{\Delta^{pw/tw}}$ makes with the coordinate axes after each increment, are ${\theta_\Delta}$ and ${\phi_\Delta}$ for plane waves and $\theta_{\Delta^{tw}}$ and $\phi_{\Delta^{tw}}$ for twisted waves.  These are calculated for each rotation by operating the rotation matrix (Equation (36)) on $\boldsymbol{\Delta}$ and $\boldsymbol{\Delta^{tw}}$. In this paper, all cross sections are orientation-averaged.

\subsection{Total (integral) cross section (TCS)}
The total elastic cross sections for all the above cases of plane and twisted waves are obtained by integrating over all solid angles of the scattered electron. 

\begin{equation}
    \sigma^{pw/tw} (E_i) = \int \frac{d\sigma}{d\Omega} d\Omega = 2 \pi \int^{\pi}_0 d\theta_s \sin{\theta_s} \left(\frac{d\sigma}{d\Omega}\right)^{pw/tw}
\end{equation}

\section{Computational Details}

We compare the bond lengths of CO$_2$ using DFT and CCSD and then perform the remaining calculations using CCSD.  We have tested the bond lengths through geometry structure optimization calculation in GAUSSIAN (G09) \cite{g09} using DFT and CCSD. In the case of DFT, we used the hybrid density functional B3LYP \cite{becke1993density}. The C=O bond length of CO$_2$ obtained from DFT using the Pople basis set 6-311G is 2.2448 $a_0$. In the case of CCSD, using the basis set cc-pVDZ, we obtain the bond length to be 2.2156 $a_0$, and with cc-pVQZ to be 2.18506 $a_0$. The experimental value of the bond length is 2.19660 $a_0$ \cite{herzberg1966molecular}. Inferring from these, we proceed with the results of the geometry-optimized CCSD calculation using the cc-pVQZ basis set. The cc-pVQZ basis set, in particular, employs 55 basis functions for each second-row periodic table element, making it computationally expensive \cite{vspirko2011potential,pak1997coupled,sanchez2020electronegativity}. A wavefunction file (.wfn file) is obtained after the CCSD calculation. The .wfn file contains all the information on the MOs and their energies corresponding to the ground state. The .wfn file is then analyzed by Multiwfn, a code developed by Lu et al. \cite{lu2012multiwfn}. We use Multiwfn to obtain the electron charge density $\rho$ of the molecule from the MO present in the wfn. file.

To model the elastic scattering, the interaction space is considered to be a spherical region with a radius of 10 $a_0$ that encompasses the molecule. Spherical coordinate integrals in Equations (19),(23),(29),(35), and (37) are computed using the Gauss-Legendre quadrature with N = 25 integration steps for each $r,\theta,\phi$. The origin of this space is taken to be at the molecule's center of mass.

\section{Results and Discussions}

\subsection{Plane wave elastic Differential Cross Sections}

Figure 3 consists of 3 subplots of DCS plotted against the scattering angle $\theta_s$ for the plane wave electron impact on a CO$_2$ molecule for incident energies 500 eV, 1 keV, and 1.5 keV.

In all three subplots, the solid dark blue curve represents the present model. The red and black points represent the experimental results obtained by Iga et al. (1984) and Bromberg (1974), respectively \cite{iga1984elastic,bromberg1974absolute}. The error bars are insignificant, given that the DCS is on a log$_{10}$ scale and are thus not shown. The DCS for 1 keV agrees with previous experiments; however, there are some deviations in the case of 500 eV. For $E_i$ = 1.5 keV, no experimental results are available in the literature.

\onecolumngrid

\begin{figure*}[h!]
    \scalebox{0.77}{\input{pwco2.pgf}}
    \caption{\justifying Elastic DCS for plane wave electron impact as a function of scattered electron angle $\theta_s$ on CO$_2$. The incident electron energies are (a) 500 eV, (b) 1 keV, and (c) 1.5 keV. The points represented by red and black dots are the experimental data obtained by Iga et al. and Bromberg \cite{iga1984elastic,bromberg1974absolute}; solid dark blue curve shows the present model}
\end{figure*}

\twocolumngrid

\subsection{Twisted-wave elastic Differential Cross Section}
We present the results for the elastic DCS of a twisted (Bessel) electron beam on CO$_2$ as a function of the scattering angle $\theta_s$. These results are presented in two categories. In the first category, angular distributions of DCS are presented at impact parameter $\textbf{b}$ = 0. In the second category, the (DCS)$_{av}$  is presented (see Equation (35)). 

\subsubsection{Differential Cross Section for $m_l$ values at $\mathbf{b} = 0$}

The DCSs plotted here are for the topological charges $m_l = $ 1, 2, 3, 7, 12, and 20 of the twisted beam. The DCS is calculated at the impact parameter $\mathbf{b}$ = 0. This scenario (or any $\mathbf{b} \neq$ 0) is difficult to realize experimentally; however, we consider this case as it gives a general understanding of the twisted electron's interaction with the target when the target lies on the beam axis. Hence, this case is purely an academic study. The opening angles used for these calculations are $\theta_p = $ 10$^{\circ}$ and $\theta_p$ = 25$^{\circ}$.

Figure 4 consists of 3 graphs of DCS plotted against the scattering angle $\theta_s$ for $m_l$ = 1, 2, and 3 of the twisted electron beam for the opening angle $\theta_p$ = 10$^{\circ}$ and energies (a) E$_i$ = 500 eV, (b) E$_i$ = 1 keV, and (c) E$_i$ = 1.5 keV. Plane-wave DCSs are also plotted (solid dark blue curve) with the twisted electron beam DCS for comparison. Twisted wave DCS for various $m_l$ values are shown in lighter colors; for $m_l = $ 1 light blue, $m_l = $ 2 pale orange, and $m_l = $ 3 light green. 

\onecolumngrid

\begin{figure*}[h!]
    \scalebox{0.77}{\input{mco2.pgf}}
    \caption{\justifying Elastic differential cross-section of CO$_2$ molecule by Bessel beam impact at $\theta_p$ = 10$^{\circ}$ with energies (a) 500 eV, (b) 1 keV, and (c) 1.5 keV for various $m_l$ at $\mathbf{b} = 0$. The legends show the various calculations performed.}
    \label{fig:enter-label44}

\end{figure*}
\twocolumngrid

From the graphs, we observe that the DCS peaks when the scattering angle $\theta_s$ equals the opening angle $\theta_p$ of the twisted beam in all cases. The presence of this peak is strikingly different from the plane wave case. This is in agreement with previous studies on twisted electrons on atomic systems \cite{kosheleva2018elastic}. The peak at $\theta_s = \theta_p$ can be directly attributed to the maximum of the intensity of the twisted electron beam at the opening angle $\theta_p$. The superposed plane waves that make up the twisted beam traverse along the surface of the cone outwards, making an angle $\theta_p$ with the beam axis. 

On comparing the DCSs of the twisted waves with the plane wave, we find that the DCS is much lower than the plane-wave impact except around a narrow region around $\theta_s = \theta_p$. This again can be inferred by the magnitude of the Bessel beam's intensity approaching zero on the beam axis where $\mathbf{b}$ = 0. The peak amplitude at 500 eV is 229 $a_0^2$, at 1 keV is 68 $a_0^2$ and at 1.5 keV is 34 $a_0^2$. There is a reduction in peak amplitude as E$_i$ is increased from 500 eV to 1.5 keV, since the DCS is inversely proportional to the fourth power of the momentum transfer, which depends on the incident energy.

The peak also narrows slightly as the energy is increased from 500 eV to 1.5 keV. Qualitatively, it is understood as: As the energy increases, $\varkappa$ present in the argument of the Bessel function increases, leading to the compression of the Bessel function. The maximum of the Bessel function narrows, pushing the scattered electrons into a narrower solid angle. The magnitude of DCS is the smallest for $m_l $ = 3 and the highest for $m_l$ = 1. 

The peak amplitude at $\theta_s = $ 10$^{\circ}$ for each energy slightly decreases as $m_l$ increases. This is due to the magnitude of the Bessel function decreasing for higher $m_l$. We also observe that as the incident energy is increased from 500 eV to 1.5 keV, the DCSs beyond $\theta_s = $ 30$^{\circ}$ seem to merge. They cross over for higher $m_l$. The crossover is a feature observed at all energies. We show it in Figure 5.

There is a smooth decrease in the angular distribution of DCS for twisted waves reported by Kosheleva et al. for atomic targets \cite{kosheleva2018elastic}. We, however, obtain slightly bumpy curves (not very smooth) that arise due to the interference of the incident beam by the multinuclear nature of molecules. Such a kind of interference occurs in the case of multinuclear molecules, which has been reported previously \cite{blanco2015interference,dogan2013double}.

Next, we present the DCS for twisted wave impact for $\theta_p$ = 25$^{\circ}$ for higher values of $m_l$ at E$_i$ = 1.5 keV. We have calculated the DCS of CO$_2$ by twisted wave impact for $m_l$ = 1 to $m_l = $ 20 with an interval of 1. The primary observation from our results is that the peak amplitude decreases in magnitude and peak width decreases around $\theta_s = \theta_p$ (25$^\circ$) as $m_l$ is increased from 1 to 20. However, for scattering angles $\theta_s$ away from $\theta_p$, the magnitudes of the DCS oscillate with $m_l$. In the region, where $\theta_s$ is far from $\theta_p$, the highest value of DCS obtained is for $m_l$ = 17 and the lowest for $m_l$ = 6. The DCS for other values lies within this range. In Figure 5, we have shown the DCS only for $m_l$ = 3, 7, 12 and 20 for brevity.

\onecolumngrid

\begin{figure*}[h!]
    \scalebox{1.155}{\input{higher.pgf}}
    \caption{\justifying Differential cross section of Bessel beam impact on CO$_2$ molecule at energy 1.5 keV and $\theta_p$ = 25$^{\circ}$ for various $m_l$ at $\mathbf{b} =0$. The legend shows the various calculations performed.}
    \label{fig:enter-label43}
\end{figure*}

\twocolumngrid

We notice that there is a broadening of the peak as $\theta_p$ is increased from 10$^{\circ}$ to 25$^{\circ}$. Particularly, it can be seen for $m_l$ = 3 in Figure 5, with that of Figure 4 (c). This widening of the peak is inferred as: When $\theta_p$ increases at a fixed energy, ${k}_{i_\parallel}$ decreases, and $\varkappa$ increases. This results in a decrease in the magnitude of the DCS and an increase in its width since the scattering amplitude in Equation (29) depends on $\varkappa$ and ${k_{i_\parallel}}$.

For this case, clearly our studies show that the variation in magnitude of DCS away from the peak oscillates with the $m_l$ values. The peaks lower in magnitude and sharpen as $m_l$ is increased.

\subsubsection{Differential cross sections  averaging over $\mathbf{b}$}

Here, we present the results of impact parameter averaged DCS ((DCS)$_{av}$). Figure 6 illustrates 3 subplots of (DCS)$_{av}$ plotted against the scattering angle $\theta_s$ at incident energies E$_i$ = 500 eV, 1 keV, and 1.5 keV. The opening angles considered for the Bessel beam are $\theta_p$ = 6$^{\circ}$, 20$^{\circ}$ and 45$^{\circ}$. They are represented by broken curves in magenta, light blue, and green, for $\theta_p$ = 6$^{\circ}$, 20$^{\circ}$, and 45$^{\circ}$, respectively. Plane-wave results are also shown for comparison (solid dark blue curve).

We observe peaks in the angular profile of the (DCS)$_{av}$ at $\theta_s$ = $\theta_p$ for all energies. This is consistent with the previous study on twisted electrons on atomic systems \cite{kosheleva2018elastic}. As $\theta_p$ increases from 6$^{\circ}$ to 45$^{\circ}$, the peak value decreases by almost two orders of magnitude between $\theta_p$ = 6$^{\circ}$ and 20$^{\circ}$ and by almost one order of magnitude between those of 20$^{\circ}$ and 45$^{\circ}$ at each incident energy (see Figure 6: (a), (b), and (c)). The decrease is explained in the discussion of Figure 5, when the results of Figures 4 and 5 were compared, for $\theta_p$ = 10$^{\circ}$ and 25$^{\circ}$ respectively.

\onecolumngrid

\begin{figure*}[h!]

    \scalebox{0.77}{\input{avco2.pgf}}
    \caption{\justifying Differential cross section of macroscopic target of CO$_2$ molecules by Bessel beam impact at energies (a) 500 eV, (b) 1 keV, and (c) 1.5 keV. The legends show the various calculations performed.}
    \label{fig:enter-label55}

\end{figure*}
\newpage

\twocolumngrid

However, as the incident energy increases from $E_i$ = 500 eV to 1.5 keV, the amplitudes of the corresponding peaks in Figure 6 (a), (b), and (c) gradually decrease for a given $\theta_p$. The decrease is nearly about one order of magnitude for all $\theta_p$. This decrease in magnitude and the broadening are also the same as in the case where $\mathbf{b}$ = 0 and are also explained in the previous subsection. 

Our study clearly shows that the peaks in the $(DCS)_{av}$ for molecular systems decrease rapidly as $\theta_s$ is increased, as opposed to the atomic case, where the $(DCS)_{av}$ gradually decreased \cite{kosheleva2018elastic}. 

\subsection{Twisted wave Total elastic Cross Section}

The twisted electron total elastic cross-sections (TCS) of CO$_2$ are presented in this section. Figure 7 consists of two subplots: (a) illustrates the TCS with $\mathbf{b}$ = 0 and $\theta_p$ = 20$^{\circ}$ for different $m_l$ values, and (b) illustrates the (TCS)$_{av}$ averaged over $\mathbf{b}$. The label on the y-axis $\sigma$ represents both TCS and TCS$_{av}$. The plane wave TCS is also plotted for reference and is shown by a continuous dark blue curve. TCS and (TCS)$_{av}$ are plotted as a function of incident energies. Our model does not accurately predict the magnitude of cross-sections at lower energies ( < 300 eV) since the Born approximation is not accurate when energies are lower (approximately below 300 eV) or when the interaction potential is too strong or both \cite{silverman1966test,russek1970modified}.

Figure 7(a) illustrates the TCS results for topological charges $m_l$ = 1, 2, and 3 at $\mathbf{b}$ = 0 and $\theta_p$ = 20$^\circ$. The TCS obtained for all three topological charges is considerably lower than the plane waves. This can be inferred from the DCS of the twisted beams being lower than that of the plane waves in Figure 4 for the case $\mathbf{b}$ = 0. As seen in Figure 7 (a), the TCS of the three topological charges decreases with energy. They range from 0.1 $a_0^2$ to 1 $a_0^2$ at $E_i$ = 1.5 keV. Figure 7 (b) illustrates the (TCS)$_{av}$ as a function of incident energy $E_i$. The (TCS)$_{av}$ follows nearly identical trends of the plane wave TCS for lower $\theta_p$ (See graphs related to plane wave, $\theta_p$ = 6$^{\circ}$ and 20$^{\circ}$). This can be inferred from the DCS plots in Figure 6 of the impact parameter averaging, where the twisted wave DCSs are greater in magnitude than those of plane waves beyond $\theta_s$ = $\theta_p$. However, for $\theta_p$ = 45$^{\circ}$ (TCS)$_{av}$ deviates further from the TCS of the plane wave. The TCS does not show any peaks but shows a gradually decreasing trend.

\onecolumngrid

\begin{figure*}[h!]
    \input{TCS.pgf} 
    \caption{\justifying Total Elastic Cross-Section as a function $E_i$ for (a) various $m_l$ at $\mathbf{b} = 0 $ for $\theta_p$ of 20$^\circ$, and (b) average over impact parameters of CO$_2$ ((TCS)$_{av}$). The legends show the various calculations performed.}
\end{figure*}

\twocolumngrid

\section{Conclusions}
We have presented a theoretical methodology to calculate the elastic differential and total cross-sections of molecules, particularly the CO$_2$ molecule, by plane wave and Bessel electron beam impacts. The cross-sections are computed at high energies in the static and the first Born approximations. We have first performed the CCSD and DFT calculations and have compared the bond lengths from the results. We proceeded with the CCSD calculations and computed the electron charge density, having accounted for the electron correlations. All cross sections are orientation-averaged. We plotted the plane wave cross sections followed by the twisted wave cross sections for various OAM at impact parameter $\mathbf{b}$ = 0. Following this, we calculated the cross-sections averaged over the target's impact parameters. All twisted calculations were compared with the plane waves. The DCS shows peaks at the scattering angles equal to the opening angles of the twisted beam. The magnitude of DCS away from the peak oscillates with the OAM of the electrons $\theta_p$. The (DCS)$_{av}$ shows peaks at the opening angles and does not depend on the OAM of the electrons. Finally, the TCS and TCS$_{av}$ do not show peaks but depend on the OAM and the opening angles of the beam, as shown in Figure 7.

This article serves as a beginning to twisted electron-molecular collisions and motivation towards performing experiments on twisted electron-molecular interactions. To support this, we have obtained theoretical results using techniques such as orientation averaging and impact parameter averaging, which mimic the conditions found in the laboratory. Theoretically, this method can be extended to inelastic scattering and ionization processes on other molecular targets to investigate the electronic structure. One can investigate the scattering at low energies for molecules using other methods and include the effects of higher Born series terms for the twisted electron molecular scattering.

\section{Acknowledgments}
The authors acknowledge the Department of Science and Technology(DST), Government of India, for funding received through the grant CRG/2021/003828. The authors also acknowledge Dr. Didier Sebilleau for his valuable suggestions on this article.


\bibliography{a}

\end{document}